\documentclass{article}
\usepackage{ijcai24}

\usepackage[table]{xcolor}
\usepackage{multirow}
\usepackage{times}
\usepackage{soul}
\usepackage{url}
\usepackage[hidelinks]{hyperref}
\usepackage[utf8]{inputenc}
\usepackage[small]{caption}
\usepackage{graphicx}
\usepackage{amsmath}
\usepackage{amsthm}
\usepackage{booktabs}
\usepackage{algorithm}
\usepackage{algorithmic}
\usepackage[switch]{lineno}
\usepackage{makecell}
\usepackage[export]{adjustbox}

\title{FAR-Trans: An Investment Dataset for Financial Asset Recommendation}

\author{
Javier Sanz-Cruzado$^1$
\and
Nikolaos Droukas$^2$\and
Richard McCreadie$^{1}$\\
\affiliations
$^1$University of Glasgow\\
$^2$National Bank of Greece\\
\emails
javier.sanz-cruzadopuig@glasgow.ac.uk,
droukas.nikolaos@nbg.gr,
richard.mccreadie@glasgow.ac.uk
}

\begin{document}

\maketitle

\begin{abstract}
    Financial asset recommendation (FAR) is a sub-domain of recommender systems which identifies useful financial securities for investors, with the expectation that they will invest capital on the recommended assets. FAR solutions analyse and learn from multiple data sources, including time series pricing data, customer profile information and expectations, as well as past investments. However, most models have been developed over proprietary datasets, making a comparison over a common benchmark impossible. In this paper, we aim to solve this problem by introducing FAR-Trans, the first public dataset for FAR, containing pricing information and retail investor transactions acquired from a large European financial institution. We also provide a bench-marking comparison between eleven FAR algorithms over the data for use as future baselines. The dataset can be downloaded from \url{https://doi.org/10.5525/gla.researchdata.1658}.
\end{abstract}

\section{Introduction}

Recent advances in the automated analysis of financial content and artificial intelligence are driving a digital transformation of financial services. These technologies represent an opportunity for banks, fund operators and fintech companies to improve their business processes, improve the quality of their decisions, and increase financial inclusion~\cite{Soldatos2022}. The investment advice sector has been disrupted by these technologies, resulting in a transition from customers only receiving assistance from certified financial advisors to new scenarios on which the advisors' decisions are supported by automated systems or where customers are directly served by robo-advisors.

Financial asset recommendation (FAR) lies at the core of these automated advice tools. For an investor, FAR identifies a list of financial securities or assets (such as stocks, bonds or funds) ranked by their suitability for the customer. This suitability is not only driven by the investor, but also by external factors like asset returns, currency value and inflation. FAR methods also need to consider the personal situation, needs and preferences of the user, represented by past investment transactions and explicit customer information (e.g. risk tolerance and investment capacity). Therefore, effective recommendations should model pricing data to distinguish highly performing and under-valued assets, while also identifying those assets on which the investor might be interested \cite{McCreadie2022,SanzCruzado2022}.

There is growing interest into the research and development of FAR technologies, as demonstrated by workshops in prominent conferences like RecSys~\cite{Bogers2022}, ICAIF\footnote{\url{https://sites.google.com/view/ml-for-investor-recsys}} and IJCAI\footnote{\url{https://sites.google.com/view/fin-recsys2024/}}. However, the absence of public datasets with realistic customer transactions that can be used to train and evaluate approaches under a common benchmark is a significant barrier to research. A majority of past research has focused on the study of profitability prediction methods~\cite{Paranjape2013,Schumaker2009,Sehgal2007,Song2017,Zheng2020} as asset pricing data is freely available. However, these approaches are fundamentally limited by their inability to model the customer. Although some works do consider complex investor information, such as customer profiles or transactions~\cite{Barreau2020,Chalidabhongse2006,Gonzales2022,Lee2014,Musto2015,Takayanagi2023,Zhao2015}, they rely on proprietary or simulated datasets that are not publicly available. 

This work aims to solve this limitation by proposing a novel dataset for the FAR task, provided by a large European financial institution. As far as we are aware, this dataset represents the first dataset in this domain containing pricing time series for multiple asset types (stocks, bonds and mutual funds), asset descriptions, as well as most importantly (anonymised) customer information and investment transactions. In this paper, we provide a description of the dataset, recommended experimental setup and an initial comparison of 11 baseline FAR approaches to support future work.

\section{Related work}

\subsection{Financial asset recommendation}

The particular nature of the financial domain has inspired a variety of recommendation techniques taking advantage of multiple data sources, such as: pricing data, investment transactions, news, social media, etc.~\cite{Zibriczky2016,McCreadie2022}. According to their main data source, we can categorize these methods in three primary groups: based on price, based on transactions and hybrid models. 

\subsubsection{Price-based methods}
The first category of FAR algorithms establishes price time series as their primary source of information to identify investment opportunities. These methods, based on the prediction of the price or performance of the securities, are not personalized~\cite{Zibriczky2016}.

Most works are based on regression techniques. The simplest methods use one or several regression models (such as a Random Forest or SVM) to estimate asset profitability~\cite{SanzCruzado2022,Yang2018} based on price or technical indicators. More complex models explore similarities between the time series of multiple assets~\cite{Feng2022,Paranjape2013,Zheng2020} or incorporate information from other data sources to generate the prediction, such as news with evidence of major events~\cite{Song2017} or trader's views about assets on social media
~\cite{Sun2018,Tu2018}.

More recently, some works have addressed this problem as a stock ranking selection task where the goal is to select a list of assets maximizing some utility function (for example, the combined predicted returns). \cite{Feng2019} represents the first work in this area, using a temporal graph convolutional network to combine asset prices and knowledge graphs. More recently, \cite{Alsulmi2022} combined pricing and fundamental asset information to train multiple learning-to-rank methods~\cite{Liu2009} for selecting stocks in the Saudi market.

\subsubsection{Transaction-based methods}
Following the classic methodology of recommender systems~\cite{Ricci2022}, the second category of FAR algorithms uses investment transactions as the core data source. These methods assume that investors follow patterns, and hence past investments can be used to model customers (either individually or as groups). 

Some of these works rely only on investment logs, developing collaborative filtering approaches based on matrix factorization~\cite{Lee2014,Zhao2015}, convolutional networks~\cite{Barreau2020} or customer clustering~\cite{Gonzales2022}. Other methods incorporate other information sources. For example, \cite{Musto2014,Musto2015,Musto2015b} design investment portfolio case-based recommendations factoring in the risk aversion of customers. Meanwhile, \cite{Takayanagi2024} proposes a demographic kNN method where user similarity is computed according to personality traits. Finally, \cite{Luef2020} develop content-based methods by adding asset information like market sector or enterprise life cycle, as well as a social recommendation approach based on trust between investors.

\subsubsection{Hybrid algorithms}
The last family of algorithms~\cite{Burke2007} combines several information sources to provide recommendations. For FAR, \cite{Chalidabhongse2006} propose an adaptive model to learn from past investments, financial technical indicators and demographic data about the customers. Meanwhile, \cite{Matsatsinis2009} combine collaborative filtering and multi-criteria decision analysis to generate a utility score for equity fund recommendation. \cite{Swezey2018} rerank the output of a collaborative filtering matrix factorization approach using the weights obtained in a portfolio optimization process. Luef et al.~\cite{Luef2020} propose a hybrid method that combines both content-based and knowledge-based components. Finally, Kubota et al.~\cite{Kubota2022} leverage card transactions and mobile usage app statistics from customers to identify companies they have interacted with in the past, and recommend them to invest on their stocks.

\smallskip

As we can see, many diverse algorithms have been proposed for the FAR task. However, the lack of a common dataset and evaluation methodology makes it impossible to fairly compare approaches in terms of effectiveness for the task. Therefore, in Section~\ref{sec:eval} we provide an evaluation benchmark comparing 11 models over our new dataset, drawn from the three algorithm classes.

\subsection{Existing Recommender Systems Datasets}
The development of recommendation technologies has been assisted by the availability of public resources for researchers and practitioners. One of the earliest efforts in the area is the original MovieLens dataset released in 1997~\cite{Harper2016}, which provided customer ratings for movies. Since their original release, multiple data collections have been published for different recommendation domains, including movies and TV series~\cite{Perez2020}, music~\cite{Bertin2011}, videogames~\cite{Pathak2017}, books~\cite{Wan2018} and points of interest~\cite{Yang2015}.

However, there is not a standard dataset for developing and comparing novel approaches in the investment domain. Besides those studies using only public pricing information~\cite{Chong2017,Feng2022,Yang2018}, some works have evaluated algorithms on datasets containing customer and transaction information. However, customer and transaction information is commonly subject to privacy concerns~\cite{Thompson2021}, so these works use private datasets, collected in agreement with banks or brokerage firms~\cite{Barreau2020,Kubota2022,Gonzales2022,Takayanagi2023}.

The only exception to this is the dataset introduced in~\cite{Musto2014,Musto2015b}\footnote{\url{http://bit.ly/financialRS_data_uniba} (Accessed 19/04/2024)}. Created in agreement with ObjectWay Financial Software, this dataset is publicly accessible and collects the investment portfolios of 1,172 users between June 2011 and 2013. Besides the portfolios, it includes information about customer needs and asset types. However, it does not provide pricing information about the assets or information which can be used to identify them, preventing researchers from testing price-based approaches.

In this paper, we aim to provide a new dataset which can be used to develop and evaluate novel FAR models, either focused on profitability prediction, investment transactions, or hybrid models combining both. We provide a description of the dataset in the next section.

\section{Dataset}\label{sec:dataset}

We introduce in this work a novel dataset for financial recommendation, which we shall name FAR-Trans. As far as we are aware this dataset represents the first public dataset containing both asset pricing information and investment transactions for FAR. The data has been provided by a large European financial institution, representing a snapshot of the market available to Greek investors between January 2018 and November 2022. FAR-Trans covers pricing data for stocks, bonds and mutual funds, as well as investment transaction logs (asset buy and sell actions) handled by the institution, customer, market and asset information. This section provides a description of the dataset and the acquisition and cleaning methodology. The dataset is available from \url{https://doi.org/10.5525/gla.researchdata.1658}. Table~\ref{tab:dataset} summarizes its global properties.
\begin{table*}[!t]
    \centering
    \footnotesize
    \caption{Description of the FAR-Trans dataset.}
    \label{tab:dataset}
    \begin{tabular}{lcclc}
        \toprule
        \multicolumn{2}{c}{Market data} & & \multicolumn{2}{c}{Customer data} \\
        \cmidrule{1-2} \cmidrule{4-5}
        Property & Value & & Property & Value\\
        \cmidrule{1-2} \cmidrule{4-5}
        Unique assets & 806 & & Unique customers & 29,090 \\
        Assets with investments &  321 & & Transactions (unique) & 388,049 (154,103)\\
        Unique markets & 38 & & Acquisitions (unique) & 228,913 (89,884) \\
        Price data points & 703,303 & & Sales (unique) & 159,136 (64,219)\\
        Average return (by assets, whole period) & 37.16\% & & Average return (by customers,  whole period)  & 22.89\% \\
        \% profitable assets & 54.28\% & & \% customers with profits & 54.56\% \\
        \bottomrule
    \end{tabular}
\end{table*}

\subsection{Prices} \label{sec:prices}
Prices indicate variations in the value of financial securities. Therefore, the past prices of financial assets represent an important source of information in the development and evaluation of FAR approaches. Pricing time-series have multiple uses: asset analysis through the computation of technical indicators, risk estimation, development of content-based FAR methods or evaluation according to the profitability of assets \cite{SanzCruzado2022}.

\subsubsection{Cleaning and pre-processing}
When acquiring pricing data for financial assets, it is not uncommon to find small gaps or invalid values in them caused by problems in the data collection. While these problems are realistic, they add a confounding variable for the asset analysis. Thus, we need to clean and pre-process our data to minimise the impact of these errors.

A potential source of error involves the collection of pricing information from multiple data sources or markets. We clean our dataset so every asset has, at most, a single price value at any given date. We first remove pure duplicates from our data. Then, for those assets that still have multiple values on a date (a) we remove values equal to 0 or (b) we keep the value which is closer to the price of the previous 5 days. In cases where the price time series changes trading currency mid-day, we keep the value closer to the price of the following 5 days instead (as these days use the new currency). 

We next treat major errors within the data: as investment transactions require capital exchange, we removed from our dataset those assets with closing price values equal to 0 at some point in their time series. We also observed assets with time gaps. For shorter periods, we can estimate the missing points, but, the longer the gap, the more inaccurate the estimation will be. We therefore then remove assets with large time gaps (longer than 10 days). 

Another aspect to study are sudden variations in price, as they might lead to outliers affecting what FAR algorithms learn. We consider as outliers those values where the price is increased by 10 times or loses 90\% of the value on a single day, and price reverts to its original value range on the following day. As those cases are mostly due to errors on data collection, we estimate the correct price by a moving average of the previous five days to prevent undesirable effects.

A more complex case occurs when the price never reverts to its original scale -- as this might reflect a currency change or a stock split. In the first case, we apply price transformations to ensure all time series are represented in euros. In the second case, stock splits represent corporate actions changing the number of shares on which a stock is divided. For instance, a company might divide every share into two -- causing every investor to own twice the number of shares, but every individual share halved in value. There are two types of splits: direct stock splits increase the number of shares, whereas reverse stock splits diminish it. To prevent variations in our data, we check those assets with increases or decreases of a third of their value in a single day. Then, with the assistance of Yahoo! Finance\footnote{Yahoo! Finance: \url{https://finance.yahoo.com/}}, we identify (a) whether a stock split occurred, (b) its date and (c) the split ratio. Then, for every stock split, we divide the prices previous to the split date by the split ratio -- for example, if a company performs a 2-for-1 (2:1) stock split, all prices previous to the split date are halved. Finally, we finish our cleaning process by closing the remaining gaps by applying a moving average over the previous five days.

\subsubsection{Statistics}
Figure~\ref{fig:asset_stats}(a) illustrates the average price of the assets included in our dataset. As we can observe, our data covers both bullish and bearish market periods -- including recent economic recess periods like the Covid-19 pandemic in March 2020, and the effects of the Ukraine-Russian war at the beginning of 2022. 

\subsection{Assets} \label{sec:assets}
We collect further information about the financial securities beyond the pricing data. For each of the 807 assets with pricing data, we obtain from public sources their asset type (stock, bond or mutual fund) and sub-type (for instance, bonds can be government or company bonds), their names, the market where they are traded and, where available, their sector and industry.

\begin{figure*}[!ht]
    \centering
    \small
    \begin{tabular}{ccc}
        \includegraphics[scale=0.24, valign=t]{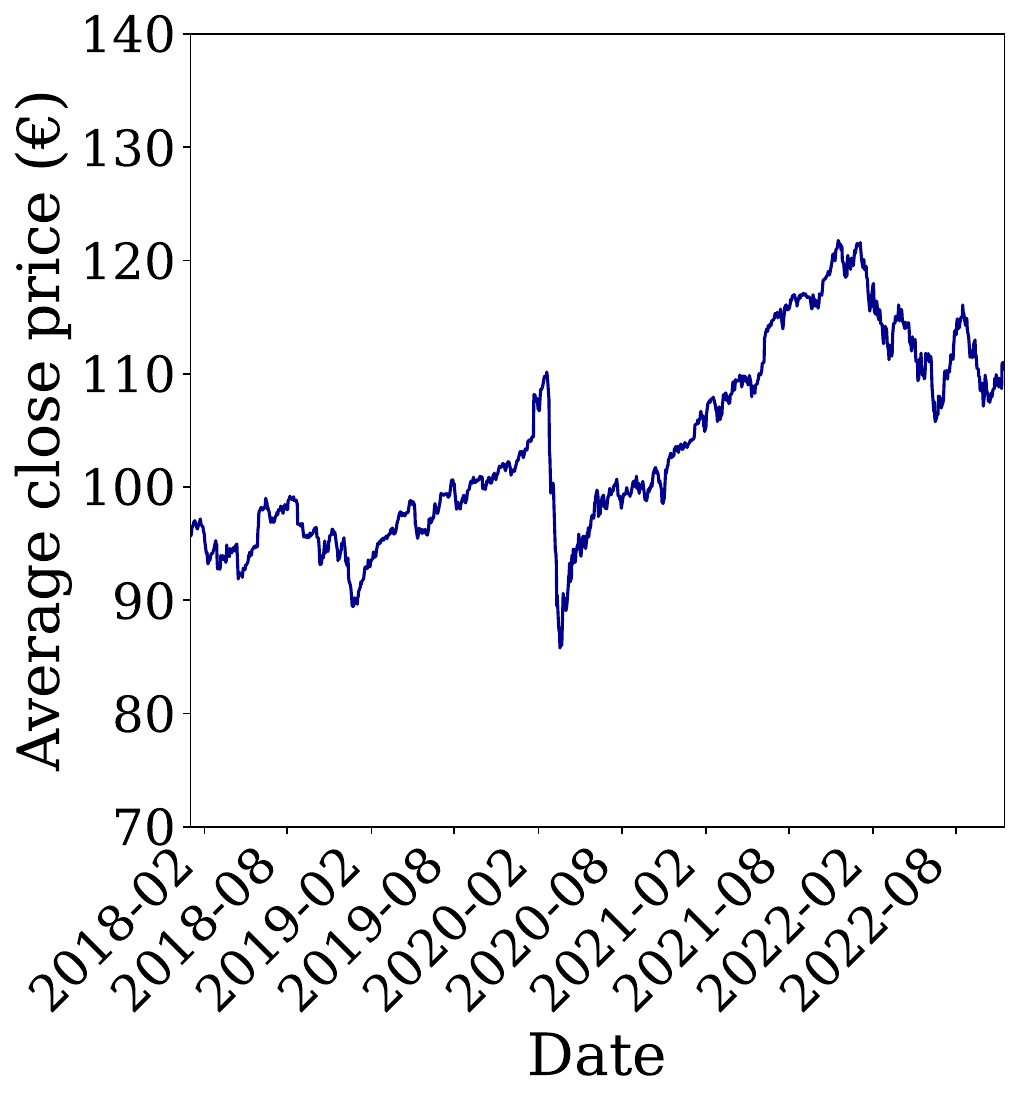} & \includegraphics[scale=0.24, valign=t]{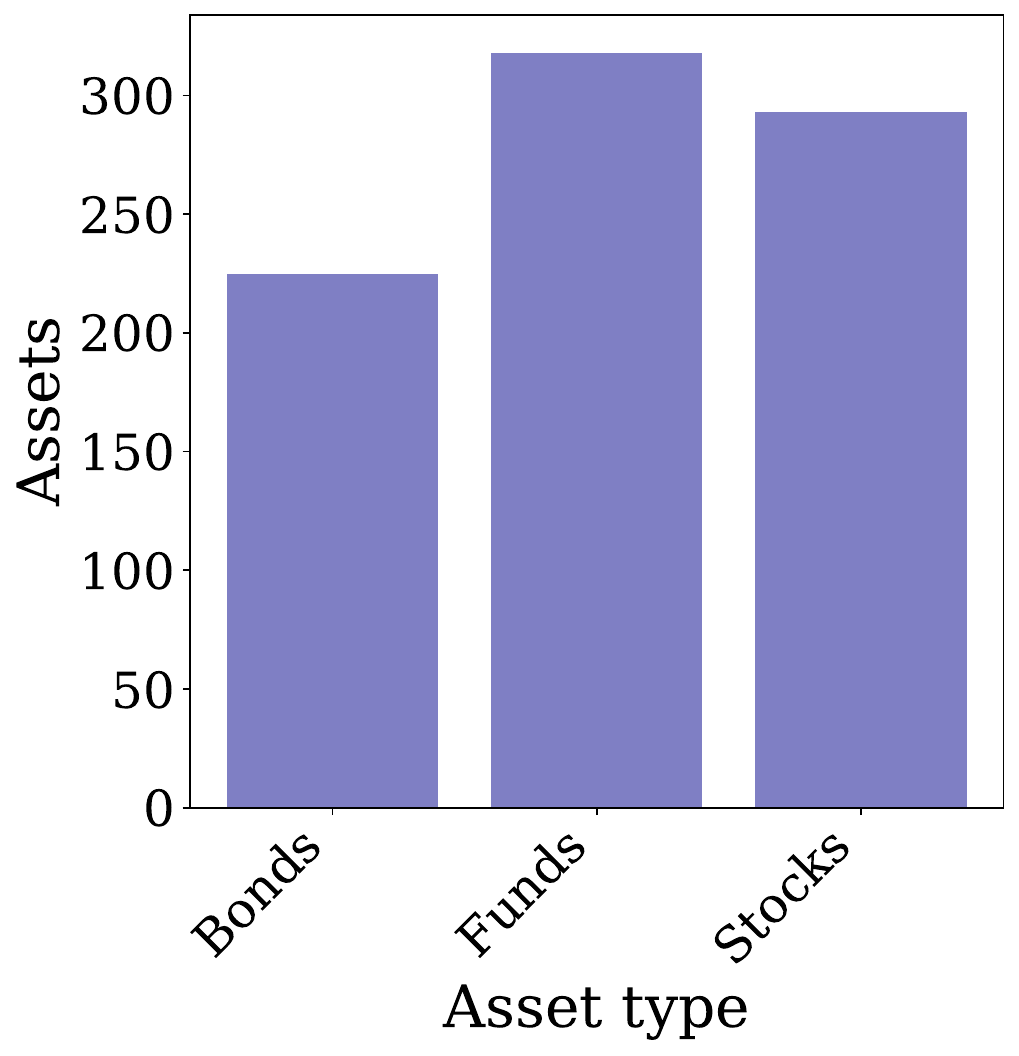} &
        \includegraphics[scale=0.24, valign=t]{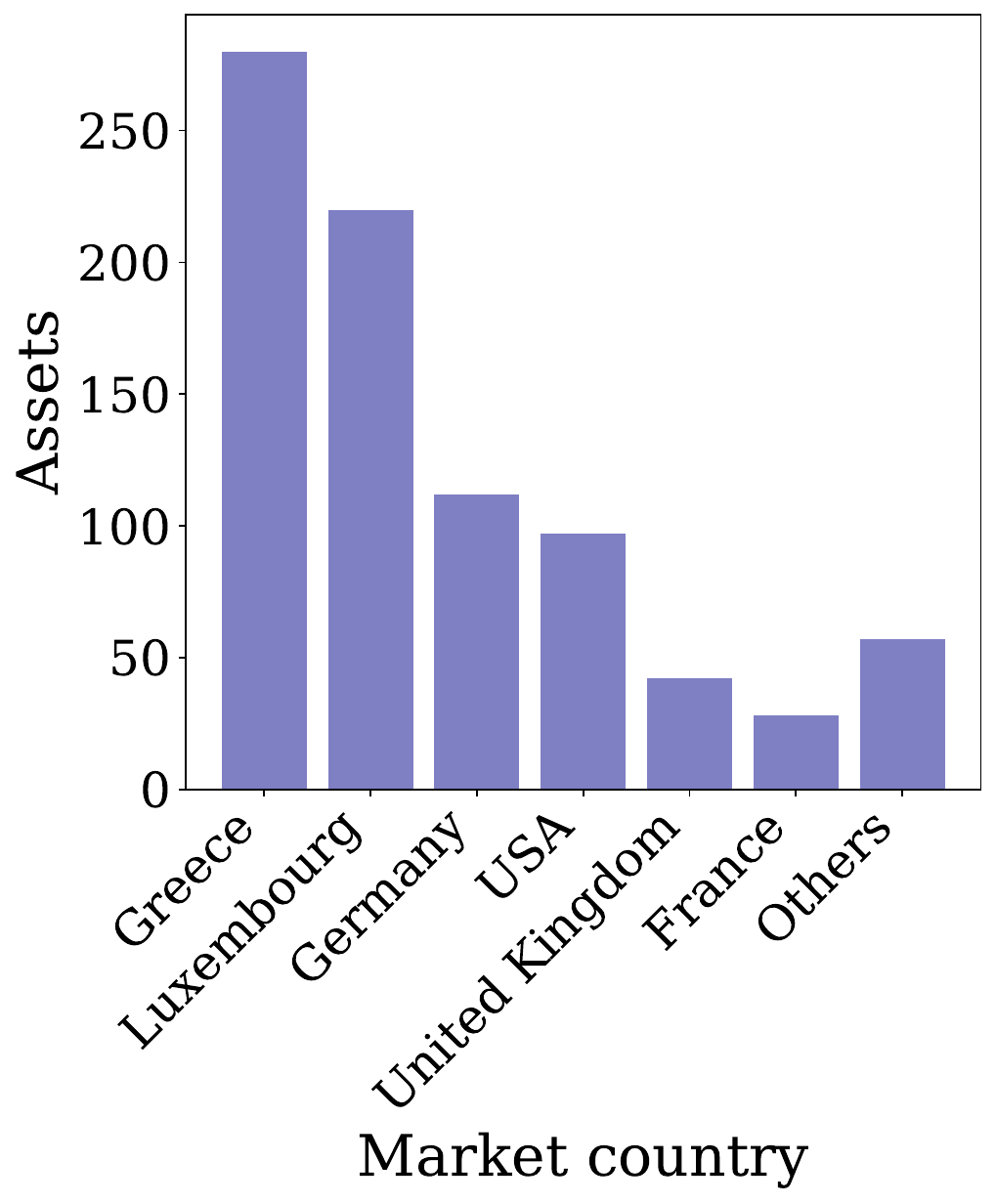} \\
        (a) Average price & (b) Asset classification by type & (c) Asset classification by market country
    \end{tabular}
    \caption{Financial asset statistics.}
    \label{fig:asset_stats}
\end{figure*}

Figures~\ref{fig:asset_stats}(b) and (c) provide some statistics about the assets. Figure~\ref{fig:asset_stats}(b) illustrates the number of assets of each type (stocks, bonds and mutual funds). As it can be observed, the three categories are well represented, with at least 200 assets on each of them -- with mutual funds representing the majority of the collection. Following Figure~\ref{fig:asset_stats}(c), we also observe that our assets are not just restricted to the Greek market -- they represent the assets on which customers could invest through the financial institution. Therefore, although a large fraction of the assets come from Greek markets, there are also assets from other European markets (e.g. Luxembourg and Germany) and some US securities.
 
\subsection{Transactions} \label{sec:trx}

The main novelty of this dataset is the availability of investment interactions between banking customers and financial assets. These interactions represent acquisitions and sales of the securities by individual investors which have been managed by the financial institution. Investment transactions can be used for modelling the past behaviour of customers and develop personalized recommendation approaches. They can also be used to evaluate investment prediction using classical recommender systems and information retrieval metrics, which consider these algorithms as a predictor for future customer behaviour \cite{SanzCruzado2022}.

\subsubsection{Cleaning and pre-processing}
The raw data includes customer and asset identifiers, the type (buy, sale) and date of the transaction, the number of shares bought/sold by the investor, the total amount of money involved in the transaction, and the channel the customer used to execute that transaction. However, this raw data needs to be aligned with the rest of the dataset (and, specifically, with the price time series described in Section \ref{sec:prices}). We therefore perform some cleaning transformations over the initial log.

First, every transaction needs an associated customer to it, so we removed those with a blank customer. Once this is done, we consider the effects that stock splits have on the number of shares every customer owns by multiplying the number of acquired/sold shares by the split ratio. If a customer owns fractional shares after a reverse stock split, we assume that the company provides cash instead of those fractional shares\footnote{https://finance.yahoo.com/news/why-investors-cash-lieu-fractional-140004745.html} and we add a transaction selling those fractional shares at the date of the split.

Another observation over the raw data is that customers sell assets which they never acquired during the 2018-2022 period, indicating that they acquired them earlier and had them in their portfolio. To ensure that every asset sale is backed by a purchase, we recreate those asset buys. For every customer, we compute the number of shares she owns of every asset. If the investor owns a negative number of shares (meaning that she has sold more shares than she has bought) at the end of 2022, we add a buy transaction at the earliest point in time for which we have pricing data for the asset (in a majority of cases, 2nd January 2018). We assume that the customer acquires the number of shares which were sold in excess.

Afterwards, we fix those cases where customers interact with assets at times when the pricing data is not available. In case the transaction is outside the range of dates for which we have pricing data, we move the transaction to the closest date where the price exists. Then, we check if customers have shares of an asset after the end of the pricing time series of the security. If they do, we add a transaction selling those assets. 

Finally, we solve inconsistencies on asset prices by providing an estimate of the total value of the transaction. We estimate the value by multiplying the number of shares by the closing price of the asset on the date of the transaction. In the end, we have 388,049 transactions in our dataset, corresponding to 29,090 customers.

\subsubsection{Statistics}
\begin{figure*}[!ht]
    \centering
    \small
    \begin{tabular}{ccc}
        \includegraphics[scale=0.24, valign=t]{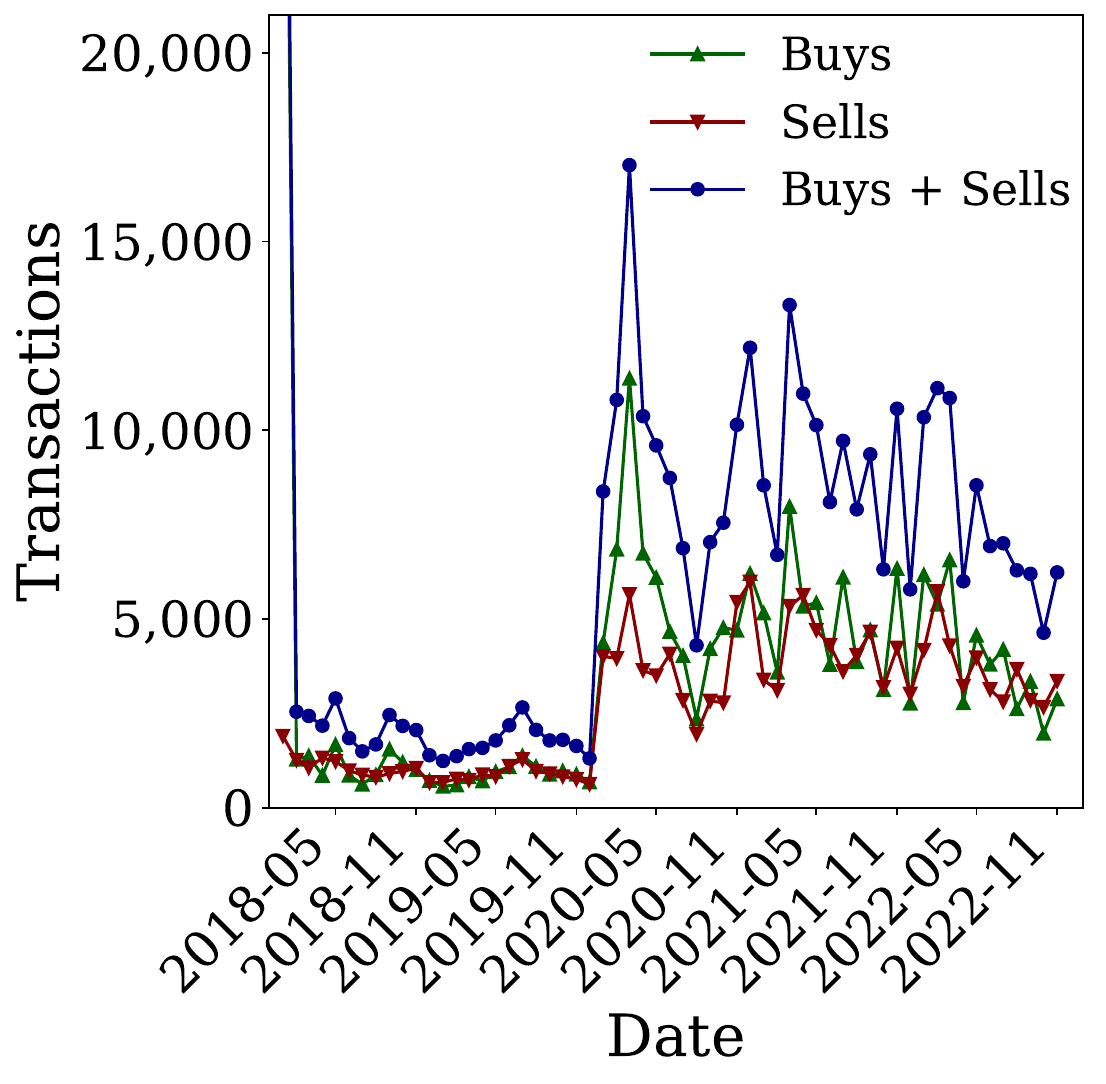} & \includegraphics[scale=0.24, valign=t]{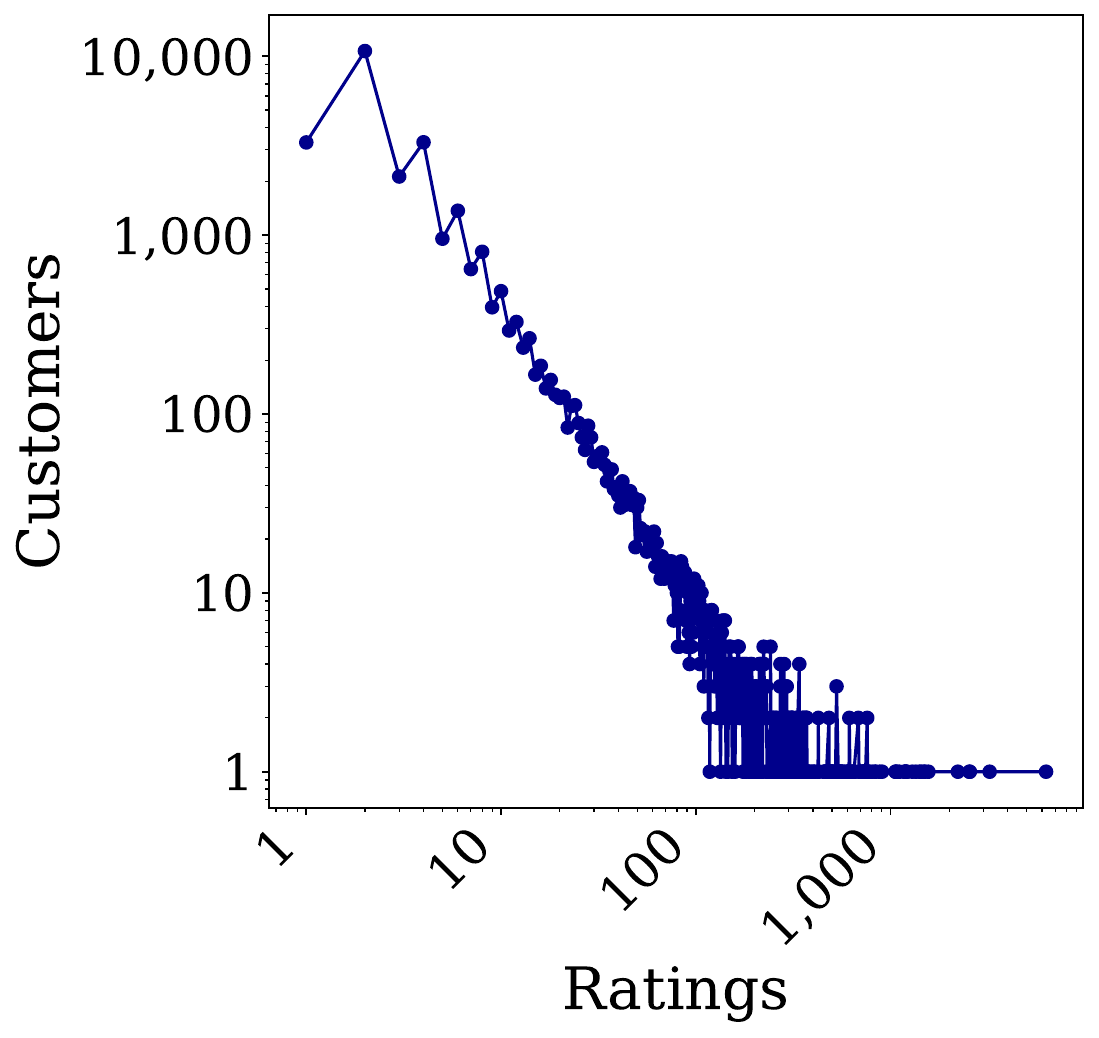} &
        \includegraphics[scale=0.24, valign=t]{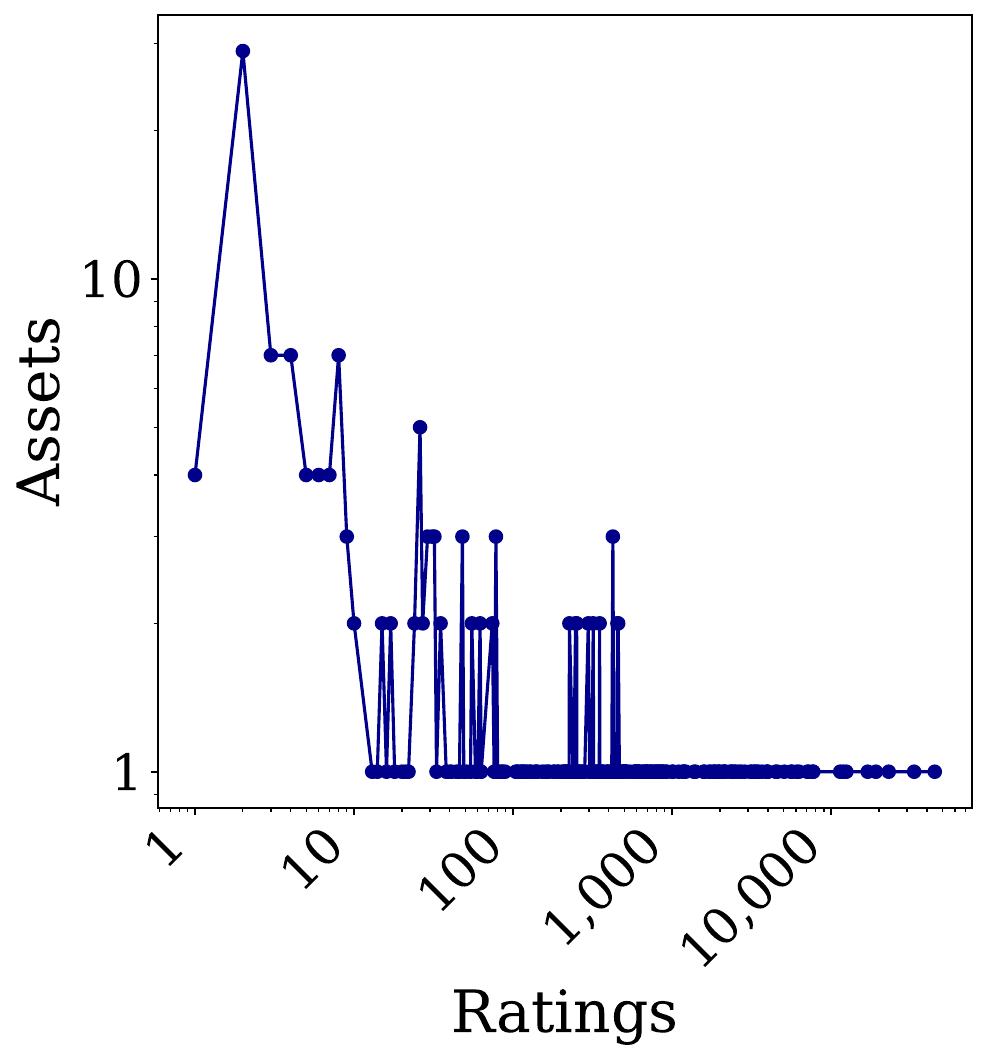} \\
        (a) Transaction distribution over time & (b) Customer transaction distribution & (c) Asset transaction distribution
    \end{tabular}
    \caption{Transaction statistics.}
    \label{fig:trx_stats}
\end{figure*}
We summarize in Figure \ref{fig:trx_stats} the statistics of the transaction data. First, \ref{fig:trx_stats}(a) displays the number of investment transactions registered on every month of the studied period. In the figure, the $x$ axis shows the dates, whereas the $y$ axis represents the number of transactions. The green line represents the asset purchases, the red line represents the sales and the blue line represents the combined number of transactions. The first observation from this figure is that most of the transactions occur in the period between January 2020 and November 2022, with over 5,000 interactions happening every month. The previous period, between January 2018 and December 2019 only receives between 1,000 and 2,000 trades per month, with the exception of January 2018. However, the large number of transactions at that date is due to the creation of asset buys representing what customers had in their portfolio before the beginning of the period covered by our dataset. The actual largest number of transactions occurs in March 2020, corresponding with the time the Covid-19 pandemic hit Europe, likely due to the huge drop in market prices that occurred during that period.

Figures \ref{fig:trx_stats}(b) and (c) represent, respectively, the transaction distribution over customers and assets. The $y$ axis represents the number of customers/assets which have associated the number of transactions as indicated on the $x$ axis. Due to the skewness of the distributions, we represent both figures in log-log scale. 

The investment distribution by customers in Figure \ref{fig:trx_stats}(b) resembles a long-tail distribution, where a majority of customers only modify their portfolios a few times over the whole period (over 50\% of the users have 3 or less transactions between 2018 and 2022) whereas only a few customers modify their investments continuously (around 650 customers have more than 100 transactions). This long-tail distribution is similar to other recommender system datasets like MovieLens.

A different pattern is observed when we examine the asset distribution however. First, as indicated in Table \ref{tab:dataset}, less than half of the assets (321 out of 807) have ever been bought or sold in our dataset. Second, although the distribution illustrated in Figure \ref{fig:trx_stats} is skewed, this is due to a few assets concentrating lots of transactions: even when most of the interacted assets have a reasonable number of transactions (75\% of them have more than 20 interactions, and 58\% of them have more than 100), the top 12 assets concentrate more than 50\% of the dataset interactions. This indicates that there is an important popularity factor over the collected transactions.

\subsection{Customers}

Financial asset recommendation needs to consider the specific needs and preferences of the customers. However, all that information is not only hidden in the past customer transactions: explicit information about customer investment capacity or risk profile can be considered to identify more relevant investment opportunities for retail customers. As such, we include in the dataset information about the classification of customers within the bank, their investment risk profile and their investment capacity. We collect that information from the 29,090 customers within the bank who have, at least, one investment reflected in our cleaned transaction data. All customer information has been thoroughly anonymized and does not contain sensitive data to satisfy regulations. We provide further descriptions for the customer classifications below:

\begin{figure*}[!ht]
    \centering
    \small
    \begin{tabular}{ccc}
        \includegraphics[scale=0.24, valign=t]{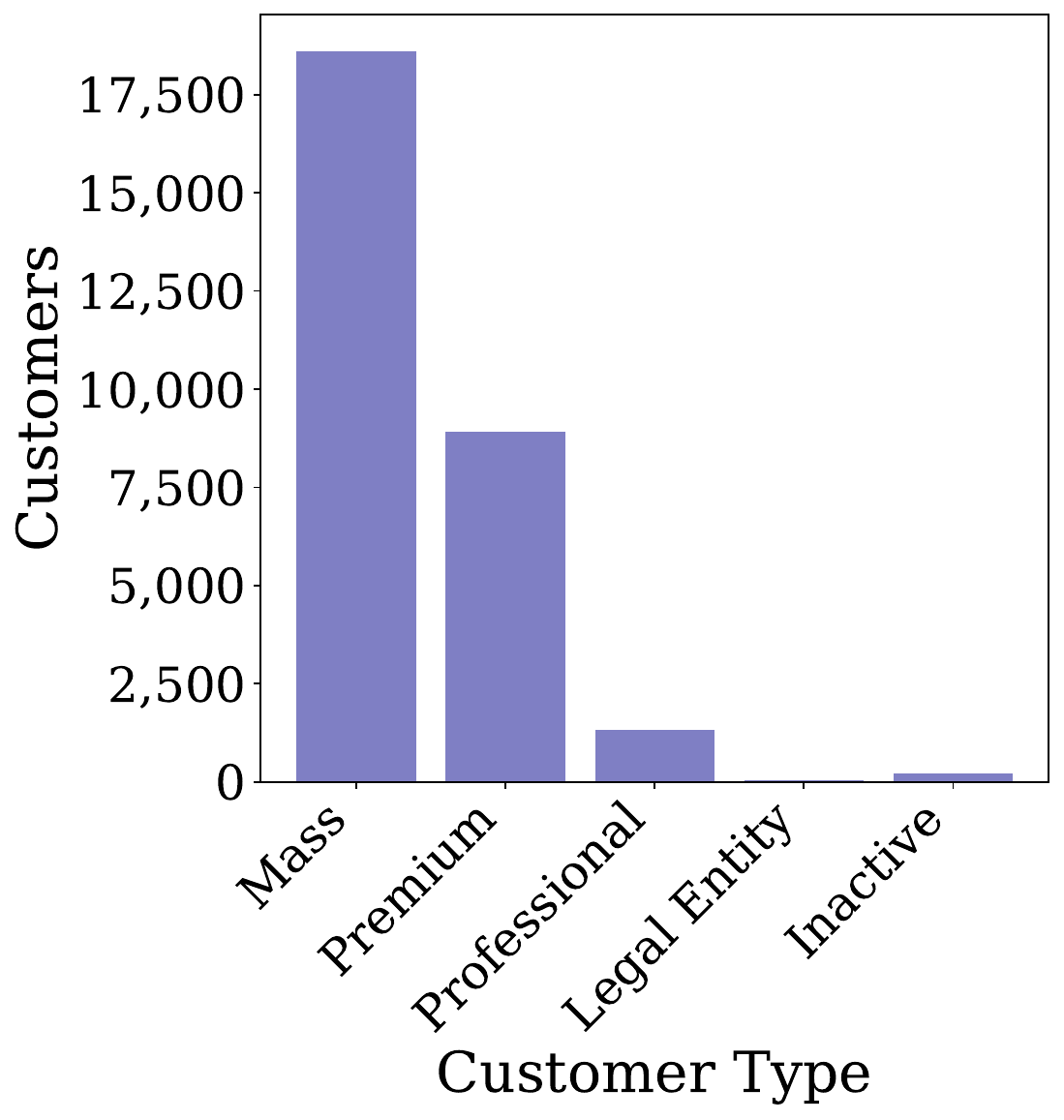} & \includegraphics[scale=0.24, valign=t]{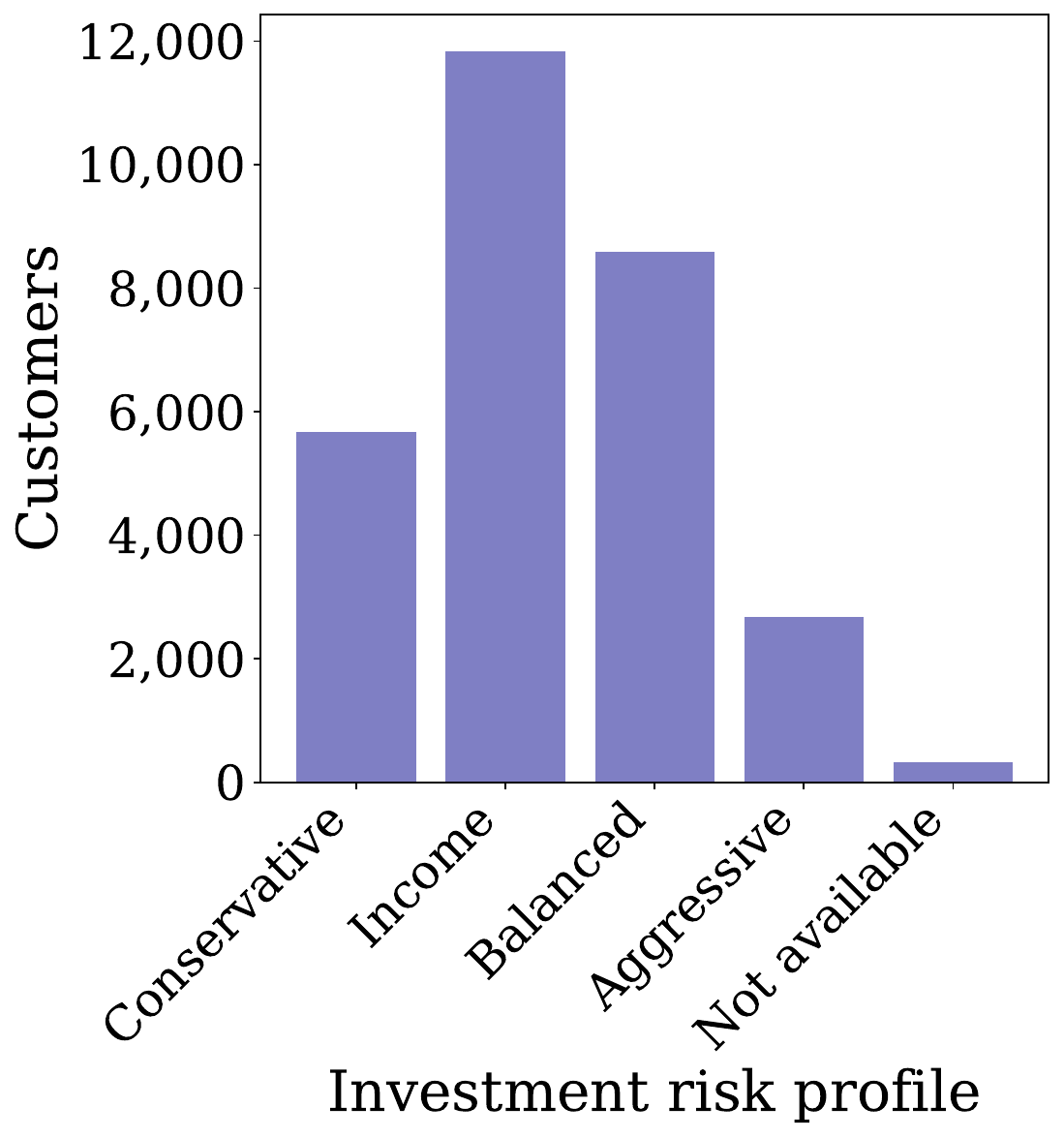} &
        \includegraphics[scale=0.24, valign=t]{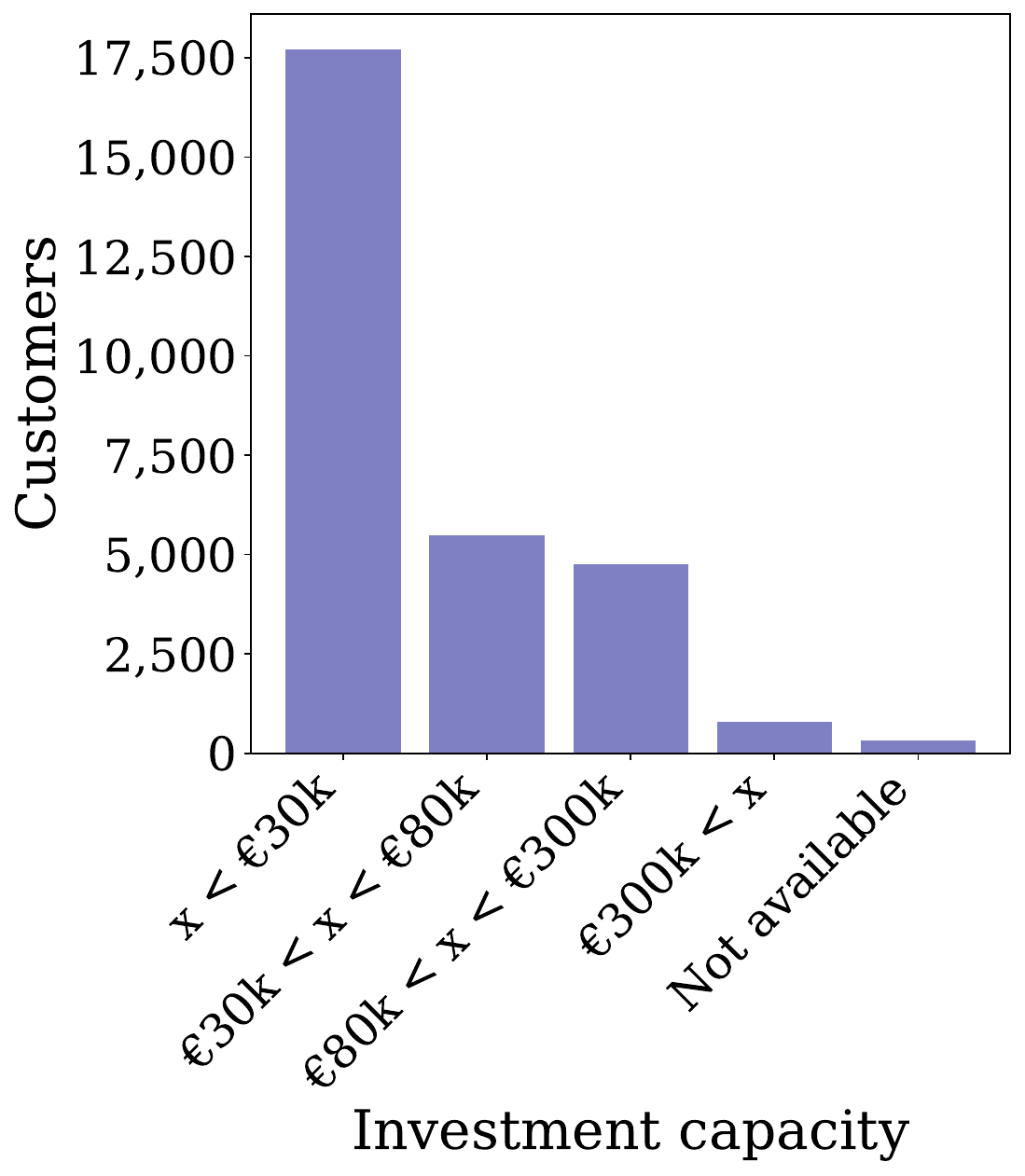} \\
        (a) Customer classification by segment & (b) Customer classification by risk profile & (c) Customer classification by capacity
    \end{tabular}
    \caption{Customer statistics.}
    \label{fig:cust_stats}
\end{figure*}

\subsubsection{Customer segments}
Customer segments represent the internal classification of customers within the bank. We consider five different segments in our data:
\begin{itemize}
    \item \textbf{Mass:} The majority of the customers. This category represents customers with less than €60,000 of managed assets (investments, deposits and insurance products).
    \item \textbf{Premium:} Individual customers with more than €60,000 on managed assets.
    \item \textbf{Professional:} Sole proprietorship. Individual customer exercising their activity without having created a legal person.
    \item \textbf{Legal Entitiy:} This category represents legal entities with services within the bank.
    \item \textbf{Inactive:} Customers without available segment.
\end{itemize}

Figure \ref{fig:cust_stats}(a) shows the distribution of customers over the different categories: as we can observe individual retail investors represent the majority of the dataset, with 18,610 mass customers, 8,906 premium customers, followed by the business customers (1,327 professional and 39 legal entities). The segment of 208 customers remains unknown.

\subsubsection{Investment risk profile}
The investment risk profile categorises customers according to the amount of risk they would accept on their investments. To assess if the offered investment assets are suitable for their investment goals and aligned with their risk aversion, every customer who is interested in investing is asked by the bank to complete an investment profile questionnaire with 25 questions. We provide these questions on the dataset. Following the MiFID II \cite{mifid2} regulatory framework, this risk assessment aims to ensure that the financial instruments provided to investors are compatible with their needs, characteristics and goals. According to their answers, their risk profile in one of the following four categories:
\begin{itemize}
    \item \textbf{Conservative:} Conservative customers prioritize protecting their capital. Their portfolios should be easily liquidated and present extremely low investment risk. An example portfolio might include short-term placements and fixed-income securities.
    \item \textbf{Income:} Customers with this risk profile aim at generating a fixed income arising from bond coupons, dividends and short-term placements. Their portfolios should present very low investment risk.
    \item \textbf{Balanced:} Accepting possible fluctuations on the invested capital, these customers aim at generating fixed income from coupons and dividends, as well as medium-term capital gains. An example portfolio would contain both bonds and stocks.
    \item \textbf{Aggressive: } This profile aims at significant long-term gains, which come with high risks.  
\end{itemize}

The risk profile is precise for those customers who have already answered the questionnaire. In case they have not, the score is simulated through an automated process: first, given the answered questionnaires, linear regression is used to  determine which questions have the largest weight. Then, estimations for the most important scores are obtained from alternative customer data (e.g., yearly salary estimation). Using estimations on those basic components and their weights, it is possible to simulate the risk tolerance for those customers without questionnaires. A risk profile is estimated following this method for 7,141 investors in our dataset.

Figure \ref{fig:cust_stats}(b) displays the distribution of customers according to their investment risk profiling. We find that most customers favor intermediate risk profiles (income and balanced), with a minority willing to risk their capital on aggressive investments. 

\subsubsection{Investment capacity}
The last customer categorization divides customers by the amount of money they can invest: we consider four different segments according to their investment limit: (a) less than €30,000, (b) between €30,000 and €80,000, (b) between €80,000 and €300,000 and (d) more than €300,000. These values are obtained from the risk assessment questionnaire. A similar procedure to the one used for risk profile is used for those customers without assessment (in this case, this is estimated for 7,318 customers).

The customer distribution is illustrated in Figure \ref{fig:cust_stats}(c). In that figure, we can observe that a majority of the investors in our dataset have a low investment capacity (less than €30k). This is consistent with our customer segmentation, where more than 18,000 customers were identified as mass customers, with less than €60k on investments. The  number of customers in each category diminishes as the investment capacity increases (with those customers capable of investments above €300k representing a minority of the dataset). 

\section{Potential use cases}
Considering the information included in the FAR-Trans dataset, we envision several potential use cases for researchers in the recommender systems and investment spaces. These use cases include (but are not limited to): 

\begin{itemize}
    \item \textbf{Investor modelling:} the customer information and investment transactions might be useful to develop new models of investor behaviour for banking customers~\cite{Thompson2021}
    \item \textbf{Financial asset recommendation:} FAR represents the main use case for which the dataset was built. The availability of customer and asset information, pricing data and transactions allow the development of price-based, transaction-based and hybrid models for the task~\cite{McCreadie2022,SanzCruzado2022}
    \item \textbf{Portfolio management:} this task involves building an investment portfolio for the customers: not only identifying investments, but also estimating how much capital they should invest on each asset and how they should modify their current investments~\cite{Markowitz1952}.
\end{itemize}

As the main use case considered during the construction of this dataset, we provide a recommended experimental setup for the FAR task, as well as algorithmic benchmarks for future comparison. 

\section{Example Use Case: Financial Asset Recommendation} \label{sec:eval}

We provide an example use case for this dataset, where we identify potential investments for retail investors using FAR algorithms. This example provides a benchmark for assessing new developments in the FAR domain. In this work, we aim to answer the following research questions:
\begin{itemize}
    \item \textbf{RQ1:} Which algorithms are best at identifying profitable assets for investors?
    \item \textbf{RQ2:} Which algorithms are best at identifying future customer investments?
\end{itemize}

\subsection{Task definition}
FAR systems consider two types of entities: investors (denoted as $\mathcal{U}$) and financial securities (denoted as $\mathcal{I}$). At a given time $t$, customers buy or sell financial assets at a price that varies according to the asset supply and demand. If we define the set of assets which a customer $u$ has bought before time $t$ as $\mathcal{I}_u(t) \subset \mathcal{I}$, a FAR system generates a ranking $R_u \subset \mathcal{I}\setminus \mathcal{I}_u(t)$ of those assets who the user has not interacted with in the past, based on their suitability for the customer.

\subsection{Experimental setup}
\subsubsection{Dataset post-processing}
\looseness -1 We modify our transaction data so our algorithms can receive it as input. We transform them into a binary rating matrix $Rel$ where every user-item pair represents the interest of the user on the item. Following common practice in implicit recommender systems \cite{Ricci2022}, we consider that a customer $u$ has interest on a financial asset $i$ ($Rel(u,i) = 1.0$) if she has acquired instances of the asset. Otherwise, it is considered that the customer is not interested in that product ($Rel(u,i) = 0.0$).

Then, as the effectiveness of different recommendation algorithms naturally varies as market conditions change \cite{SanzCruzado2022}, it is important to examine the performance over different market conditions. To this end, we generate 61 distinct variants of the dataset, each representing a setting for a different point in time. Each variant defines a time $t$ when recommendations are provided, and takes pricing data and investment transactions prior to $t$ as the training data, and the pricing data and transactions in the following $(t,t+\Delta t)$ period as test. We choose $\Delta t$ equal to 6 months in our experiments (i.e. we predict prices/interactions 6 months into the future). Our first time point $t_0$ is August 1st 2019 (providing 1.5 years worth of training data in the first instance). Time points $t \in T$ are spaced two weeks apart, so $t_1$ is mid August, $t_2$ is the beginning of September, and so on.
To avoid contamination of the test set, if a customer acquires an asset in both training and test sets, we only keep the training interactions. We also keep only those customers with at least one interaction in the training and test sets and assets that have pricing in the complete test period. This post filtering is important, as otherwise the pricing-based metrics and transaction-based metrics would be calculated over different customer and asset subsets, which would make them non-comparable.

\subsubsection{Metrics}
Following \cite{SanzCruzado2022}, we provide results for two evaluation metrics: one measuring the profitability of the provided recommendations, and another one measuring the capacity of the model to predict customer preferences.
\begin{itemize}
    \item \textbf{ROI@k:} As a measure of profitability of the recommendations, we report the monthly average return on investment (ROI) of an equally weighted portfolio containing the top $k$ recommended assets after a fixed time $\Delta t$. This measures how much our money would increase (or decrease) every month if we invested on it at time $t$.
    \item \textbf{nDCG@k:} We measure how close the recommendations are to the investments made by customers in the $(t,t+\Delta t)$ period using the normalised cumulative discounted gain (nDCG) metric~\cite{Jarvelin2002} over the top $k$ recommendations. It prioritizes relevant assets (i.e. assets acquired during the test period) in the top ranks. We consider that an asset is relevant only if $u$ acquires $i$ during the $(t,t+\Delta t)$ period.
\end{itemize}
For both metrics, we use $k=10$ and $\Delta t$ equal to six months.

\subsubsection{Algorithms}
To provide a meaningful comparison of evaluation methods, we need to apply these methods over a range of different FAR approaches, hence, we implement a diverse suite of 11 FAR approaches from the literature, summarized below:
\begin{itemize}
    \item \textbf{Random recommendation:} As a sanity-check baseline, we include an algorithm recommending assets randomly to customers.
    \item \textbf{Profitability-based models:} We test three regression algorithms, predicting ROI at $t + 6$ months: linear regression, random forest and LightGBM regression \cite{Guolin2017}. We craft a selection of technical indicators based on close price as features: average price, return on investment, volatility, moving average convergence divergence, momentum, rate of change, relative strength index, detrended close oscillator, return on investment/volatility ratio, and maximum and minimum values over a time period prior to prediction.
    \item \textbf{Transaction-based models:} We choose several methods exploiting investment transactions to generate recommendations. We divide them in two categories:
    \begin{itemize}
        \item \textbf{Non-personalized:} As a baseline, we consider popularity-based recommendation, which ranks assets according to the number of times they have been purchased in the past.
        \item \textbf{Collaborative filtering:} As collaborative filtering methods, we test three proposals: LightGCN \cite{He2020}, matrix factorization (MF) \cite{Rendle2020} and user-based kNN (UB kNN) \cite{Nikolakopoulos2022}. We also add the Apriori association rule mining (ARM) algorithm~\cite{Agrawal1994}.
    \end{itemize}
    \item \textbf{Hybrid methods:} Finally, we test two hybrid methods, based on gradient boosting regression trees~\cite{Guolin2017,SanzCruzado2022}: a regression LightGBM algorithm, targeting the profitability at six months in the future (Hybrid-regression), and, second, the LightGBM implementation of the LambdaMART learning to rank algorithm~\cite{Burges2010}, optimizing nDCG (Hybrid-nDCG). As features, we use the outcome of all the previous listed recommendation algorithms.
\end{itemize}
For each algorithm, we select as the optimal hyperparameters those maximizing the ROI at 6 months at three dates: April 1st 2019, October 1st 2019 and January 31st 2020.

\subsection{Experimental results}

In order to foster research over this dataset, we provide a benchmark of multiple FAR models on the dataset. We therefore report the performance of  the 11 FAR approaches in Table~\ref{tab:summaryresults2years} where every column represents one evaluation metric averaged over all the considered time points. The highest performing model under each metric is highlighted in bold and underlined, and the performance distribution for each metric is colour coded (blue for highly performing and red for poorly performing). From Table~\ref{tab:summaryresults2years} we observe the following points of interest:

\begin{table}[t]
    \centering
    \footnotesize
    \caption{Effectiveness of the compared models at cutoff 10. A cell color goes from red (lower) to blue (higher values) for each metric, with the top value both underlined and highlighted in bold. For ROI, blue cells show an improvement over the average market value.}
    \label{tab:summaryresults2years}
    \begin{tabular}{lccc}
        \toprule
        Data source & Algorithm & nDCG@10 & ROI@10 \\
        \midrule
        None & Random & \cellcolor[HTML]{F8696B} 0.0106 & \cellcolor[HTML]{FBECEF} 0.0071 \\
        \midrule
        \multirow{3}{*}{Prices} &  Random forest & \cellcolor[HTML]{F87577} 0.0237 & \cellcolor[HTML]{5A8AC6} \underline{\textbf{0.0259}} \\
        & Linear regression & \cellcolor[HTML]{F87375} 0.0215 & \cellcolor[HTML]{6491CA} 0.0249 \\
        & LightGBM & \cellcolor[HTML]{F87375} 0.0221 & \cellcolor[HTML]{79A0D1} 0.0225 \\ 
        \midrule
        \multirow{5}{*}{Transactions} & Popularity & \cellcolor[HTML]{9CB9DE} 0.2710 & \cellcolor[HTML]{F86D6F} 0.0006 \\
        & LightGCN & \cellcolor[HTML]{5A8AC6} \cellcolor[HTML]{5A8AC6} \underline{\textbf{0.3404}} & \cellcolor[HTML]{F8696B} 0.0004 \\ 
        & ARM & \cellcolor[HTML]{ABC3E3} 0.2556 & \cellcolor[HTML]{F86F71} 0.0007  \\
        & MF & \cellcolor[HTML]{F4F6FC} 0.1780 & \cellcolor[HTML]{F9ABAD} 0.0038 \\ 
        & UB kNN & \cellcolor[HTML]{FBF3F6} 0.1599 & \cellcolor[HTML]{D8E3F3} 0.0119 \\ 
        \midrule
        \multirow{2}{*}{Hybrid} & Hybrid-nDCG &  \cellcolor[HTML]{C2D3EB} 0.2313 & \cellcolor[HTML]{FBDCDF} 0.0063 \\
        & Hybrid-regression & \cellcolor[HTML]{F87779} 0.0261 & \cellcolor[HTML]{CDDBEF} 0.0132 \\ 
        \midrule
        \multicolumn{2}{l}{Market average} & - & \cellcolor[HTML]{FFFFFF} 0.0079 \\ 
        \multicolumn{2}{l}{Customer average} & - & \cellcolor[HTML]{FFFFFF} 0.0018 \\ 
        \bottomrule
    \end{tabular}
\end{table}

First, we observe that, in general, only a few of the algorithms are able to provide a set of assets which are profitable above the market ($\mbox{ROI}@10 > 0.0079$): the price-based algorithms, the hybrid model optimizing a profitability regression function and the user-based kNN collaborative filtering algorithm. Among these, the best alternatives are notably the profitability prediction models, with the three of them (linear regression, random forest and LightGBM) being able to beat the monthly profitability of a market fund where all assets are equally weighted. From these three models, random forest regression appears as the best alternative. However, these methods fail to identify assets on which customers are interested (achieving nDCG values barely above random recommendation).

Second, transaction-based algorithms are able to reasonably predict customer preferences (as shown by their high nDCG values). We observe that the algorithm with the highest nDCG value is the most advanced LightGCN algorithm. However, we can also notice that the rest of the approaches tested are not able to outperform the non-personalized popularity-based recommendation algorithm. This follows our previous observation that 10 assets concentrate around 50\% of the investment transactions in our dataset. Although collaborative filtering approaches achieve high nDCG values in our comparison, they show an overall poor performance in terms of the ROI profitability metric.

These observations allow us to answer RQ1 and RQ2: \textit{those methods targeting a particular evaluation objective are the best optimizing that perspective at evaluation
time, with price-based methods like random forest or LightGBM achieving high profitability values, whereas collaborative filtering transaction-based approaches like LightGCN stand out against other algorithms in terms of nDCG.}

\section{Conclusions}

In this work, we have introduced FAR-Trans, a novel dataset for financial asset recommendation that includes customer and asset information, asset pricing time series and investment transaction data from a large financial institution. The dataset spans the period between January 2018 and November 2022, covering not only bullish periods, but also periods of time impacted by external events such as the Covid-19 pandemic or the Ukraine-Russia war. 

We also compare 11 recommendation algorithms in terms of their capability for recommending profitable assets and their capacity for predicting future customer investments. We find that the non-personalized profitability prediction algorithms are more capable of navigating the market prices and are therefore able to provide asset recommendations above the average market profitability. On the other hand, they fail at predicting customer investments, a task on which collaborative filtering models excel.

As future work, we shall explore the use of this dataset for multiple tasks, not only including financial asset recommendation, but also portfolio construction and optimisation or investor and asset modelling. 

\section*{Acknowledgments}

The work introduced in this paper was in part carried out within the Infinitech project which is supported by the European Union's Horizon 2020 Research and Innovation programme under grant agreement no. 856632. Subsequent development was also financially supported via Engineering and Physical Sciences Research Council (EPSRC) Impact Accelerator, part of UK Research and Innovation (UKRI) with grant ref. number EP/X525716/1.

\bibliographystyle{named}
\bibliography{ijcai24}

\end{document}